\documentclass[12pt, a4paper, titlepage]{article}
\usepackage{lipsum, amssymb, amsfonts, amsthm, amsbsy}
\usepackage{upgreek}
\usepackage[dvips]{graphicx}
\usepackage[pageshow]{supertabular}
\usepackage{array}
\newcolumntype{C}[1]{>{\centering\arraybackslash}p{#1}}
\usepackage[english]{babel}
\usepackage[only, tenrm]{rawfonts}
\usepackage[intlimits]{amsmath}
\usepackage{latexsym, epic, eepic, epsf}
\usepackage{color}
\usepackage{cite}
\usepackage{textcomp}
\makeatletter

\ifx\documentclass\undefined
\else
\fi

\advance\textheight by \topskip
\ifcase \@ptsize
  \if@twoside
  \else
  \fi
\or
  \if@twoside
  \else
  \fi
\or
  \if@twoside
  \else
  \fi
\fi
\def\WideMargins{%
  \newdimen\ExtraWidth
  \ifcase \@ptsize
    \ExtraWidth = 0.5in
    \@widemargins
  \or
    \ExtraWidth = 0.5in
    \@widemargins
  \or
    \ExtraWidth = 0.7in
    \@widemargins
  \fi\let\WideMargins\relax\let\@widemargins\relax}
{\def\do{\noexpand\do\noexpand}
 \xdef\@preamblecmds{\@preamblecmds \do\WideMargins}
}
\def\@widemargins{%
    \global\advance\textwidth by -\ExtraWidth
    \global\advance\marginparwidth by \ExtraWidth
    \if@twoside
      \tw@sidedwidemargins
    \else
      \@nesidedwidemargins
    \fi}
\def\tw@sidedwidemargins{%
    \if@reversemargin
      \@tempdima=\evensidemargin
      \advance\@tempdima by -\oddsidemargin
      \advance\oddsidemargin by \ExtraWidth
      \advance\oddsidemargin by \@tempdima
      \advance\evensidemargin by -\@tempdima
    \else
      \advance\evensidemargin by \ExtraWidth
    \fi}
\def\@nesidedwidemargins{%
    \if@reversemargin
      \advance\oddsidemargin by \ExtraWidth
      \advance\evensidemargin by \ExtraWidth
    \fi}
\ifx\documentclass\undefined
\else
\fi

\makeatother

\newcommand{\RM}{R_{\text M}}                           % R_M
\usepackage{subfigure}
\usepackage{supertabular}
\newlength{\Columnsize} \Columnsize8.5cm

\usepackage{setspace}
\usepackage{lineno} %\linenumbers
\begin{document}
%\makeatletter
%\renewcommand{\@makecaption}[2]{%
%\vskip\abovecaptionskip
%\hbox to \hsize{\hfil#1\hfil}%
%\vskip\belowcaptionskip}
%\makeatother

%\renewcommand\listfigurename{Figure Legends}

\begin{titlepage}
\begin{center}
%{\Large\bf An Adaptive Genetic Algorithm for Misalignment Estimation in Spiral, Sequential and Circular Cone--Beam %Micro--CT}
{\Large\bf The rotate--plus--shift C--arm trajectory:\\
Complete CT data with less than $\text{180}^\circ$ rotation}
\\[3cm]
{\large\bf 
Ludwig Ritschl$^{1}$, Jan Kuntz$^{2}$ and Marc Kachelrie\ss$^{2}$\\[1ex]
}
{
\sl
$^1$ Ziehm Imaging GmbH, Donaustra\ss e 31, 90451 N\"urnberg, Germany\\[1ex]
$^2$ Medical Physics in Radiology, German Cancer Research Center (DKFZ), Im Neuenheimer Feld 280, 69120 Heidelberg, Germany\\[1ex]
}

\end{center}
\vfill

\begin{center}
\sl\large
%\today\\[1cm]
%Submitted to Medical Physics February 2012\\
%Revised version, submitted  ...\\
%Re--revised version, submitted ...\\
%Accepted for publication ... \\
\end{center}
\vfill

\noindent Send correspondence and reprint requests to:

Dr. Ludwig Ritschl

Ziehm Imaging GmbH,
Donaustra\ss e 31,
90451 N\"urnberg, Germany

e--mail: ludwig.ritschl@ziehm-eu.com
%\\[1cm]
\end{titlepage}

%\noindent\bf Medical Physics MS\#11-973
\begin{abstract}
\textbf{Background:} 
In the last decade C--arm--based cone--beam CT became a widely used modality for intraoperative imaging. Typically a C--arm scan is performed using a circle--like trajectory around a region of interest. Therefor an angular range of at least $\text{180}^\circ$ plus fan--angle must be covered to ensure a completely sampled data set. This fact defines some constraints on the geometry and technical specifications of a C--arm system, for example a larger C radius or a smaller C opening respectively. These technical modifications are usually not benificial in terms of handling and usability of the C--arm during classical 2D applications like fluoroscopy. The method proposed in this paper relaxes the constraint of $\text{180}^\circ$ plus fan--angle rotation to acquire a complete data set. This enables for CT like 3D imaging with a wide range of C--arm devices which are mainly designed for 2D imaging.

\textbf{Methods and Results:}
The proposed C--arm trajectory requires a motorization of the orbital axis of the C and of ideally two orthogonal axis in the C plane. The trajectory consists of three parts: A rotation of the C around a defined iso--center and two translational movements parallel to the detector plane at the begin and at the end of the rotation. Combining these three parts to one trajectory enables for the acquisition of a completely sampled dataset using only  $\text{180}^\circ$ minus fan--angle of rotation.
The method is evaluated using a mobile C--arm prototype.

\textbf{Conclusions:} 
The proposed method makes 3D imaging using C--arms with less than $\text{180}^\circ$ rotation range possible. This enables for integrating full 3D functionality into a C-- arm device without any loss of handling and usability for 2D imaging. We expect that the transition of this method into clinical routine will lead to a much broader use of intraoperative 3D imaging in a wide field of clinical applications.

\end{abstract}
%\marginsize{1in}{1in}{1in}{1in}

%==========================================================================================
%==========================================================================================
\section{Introduction}
In the last decade C--arm--based cone--beam CT became a widely used modality for intraoperative imaging \cite{fahrig:00,kalender:07, strobel:09}. Depending on the application there are different realizations of those systems either based on fixed installed C--arm units or based on mobile C--arm devices. While the former are typically used in angiographic interventions the latter are mainly used in orthopedic surgery today. Over the last years, however, there is a trend to high end mobile C--arm devices which offer comparable image quality and radiation time like fix installed systems, so it is expected that fields of applications will not be separated in future as they are today.   

All those C--arm devices have in common that a scan is performed using a circle--like trajectory around a region of interest. Thereby an angular range of at least $\text{180}^\circ$ plus fan--angle must be covered to ensure a completely sampled data set \cite{parker:82}. This yields the so--called short scan trajectory which is assumed to be the gold standard for C--arm--based cone--beam CT today.  There are different realizations of the short scan trajectory, which can be separated into orbital rotatations of the C, the so--called propeller rotation (angular rotation), or completely free circle trajectories based on robotic systems. Regarding the orbital C--arm rotation, which is the mostly used, the requirement of $\text{180}^\circ$ plus fan--angle rotation range puts some constraints on the geometry and technical specifications of a C--arm system, for example a larger C radius or a smaller C opening respectively. These technical modifications are usually not beneficial in terms of handling and usability of the C--arm during classical 2D applications like fluorosopy. If C--arms with less than $\text{180}^\circ$ plus fan--angle rotation range are used for 3D imaging, the mentioned problems related to their handling do not occur, but 3D image quality is strongly affected by limited angle artifacts in the reconstructed images. In the literature different approaches for image reconstruction of datasets acquired with less than $\text{180}^\circ$ plus fan--angle rotation range can be found. They can be separated into methods which reconstruct only parts of the field of view (super short scan) \cite{noo:02b} and iterative techniques which try to compensate for the missing data using a priori knowledge of the object \cite{sidky:06,ritschl:11}. All these techniques have in common, that they do not offer a comparable image quality or a comparable size of the reconstructed volume compared to the short scan.

The method proposed in this paper relaxes the constraint of $\text{180}^\circ$ plus fan--angle rotation of the C--arm for the acquisition of a complete data set while ensuring the full angular sampling of the reconstructed volume over the full field of view. This enables for CT--like 3D imaging with a wide range of C--arm devices which are mainly designed for 2D imaging.
\section{Materials and Methods}
\subsection{Basic idea}
\begin{figure}[t!]
\centering\includegraphics[width=\textwidth]{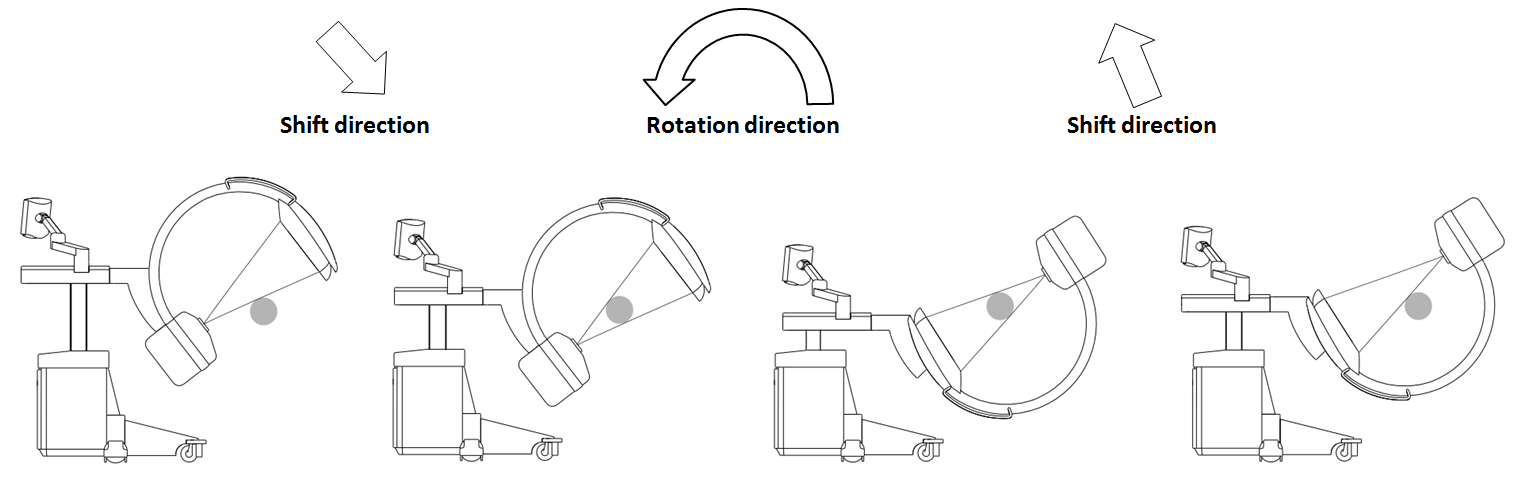}
\caption{Exemplary steps of the proposed rotate--plus--shift trajectory are shown for a robotic motorized mobile C--arm system. 
}
\label{ShiftImages}
\end{figure}

\begin{figure}[t!]
\centering\includegraphics[width=\textwidth]{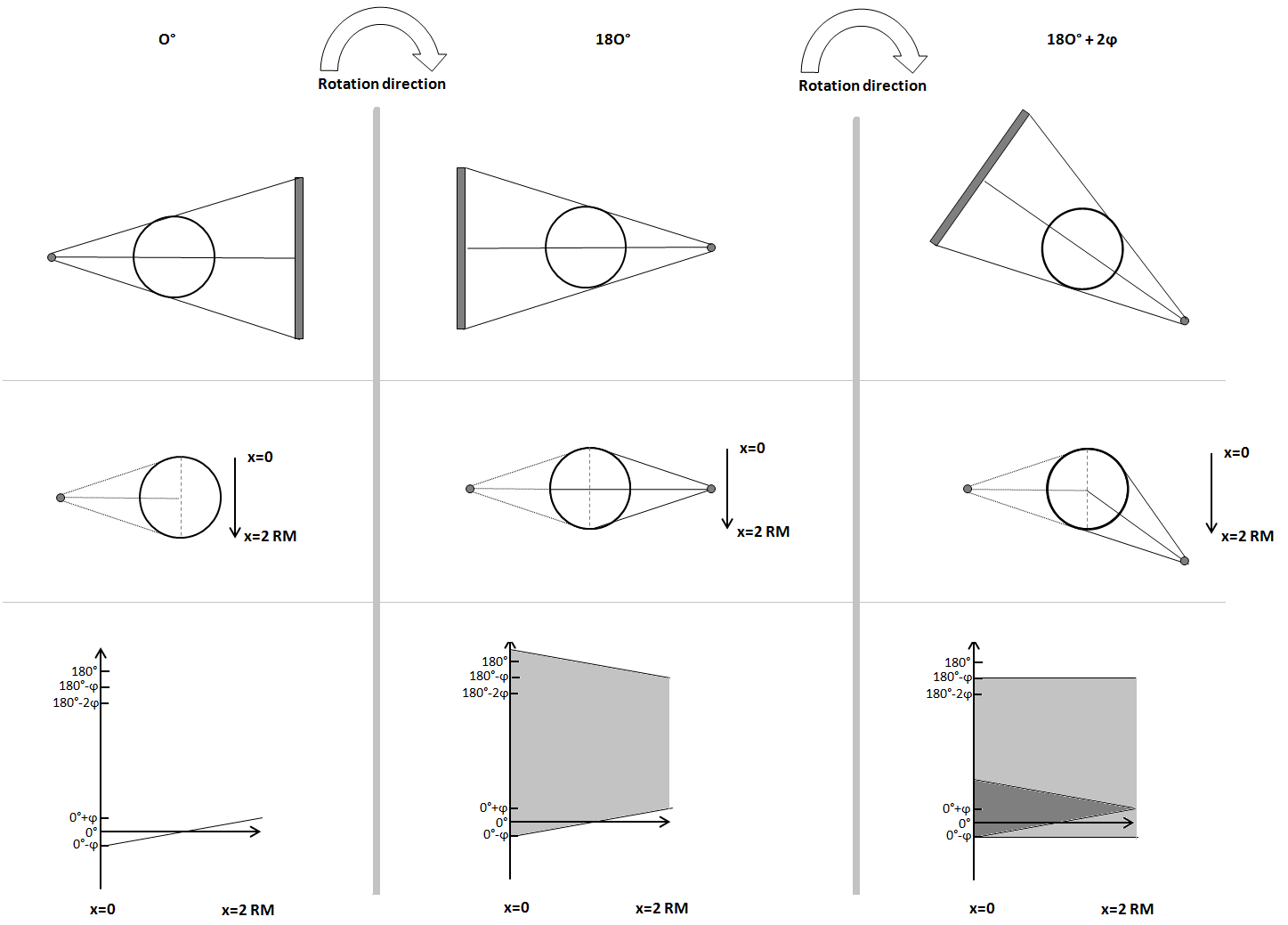}
\caption{In this figure the angular sampling in different points inside the field of measurement is shown. The used source trajectory is a short scan covering $\text{180}^\circ$ plus fan--angle angular range. The half fan angle is denoted with $\varphi$. The upper row shows the position of the x--ray source and the detector. In the middle row one can see the minimum angular position and the current angular position of the intersecting ray in three points inside the field of measurement. As one can clearly see, the condition of a $180^\circ$ sampling range in the point $x=2\RM$ requires $180^\circ+2\varphi$ of rotation. The diagram in the lower row is assumed to be periodic over a range of $180^\circ$. The dark region in the lower right diagram marks rays which were measured twice and lead to redundant data.}
\label{ShortRays}
\end{figure}

\begin{figure}[t!]
\centering\includegraphics[width=\textwidth]{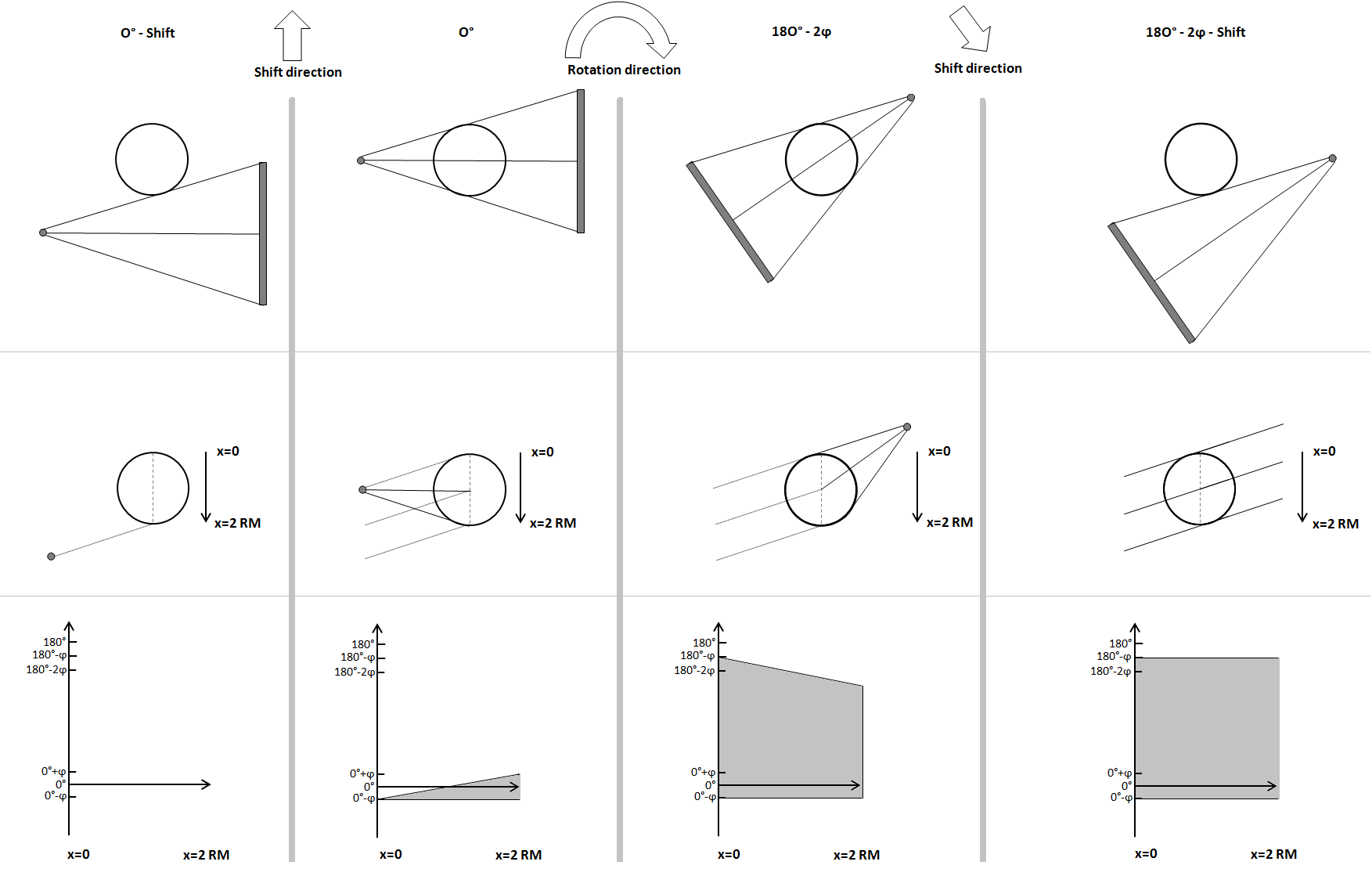}
\caption{In this figure the angular sampling in different points inside the field of measurement is shown. The used source trajectory is a rotate--plus--shift scan. The half fan angle is denoted with $\varphi$. The upper row shows the position of the x--ray source and the detector. In the middle row one can see the minimum angular position and the current angular position of the intersecting ray in three points inside the field of measurement. Note, that the condition of $180^\circ$ sampling range in the point $x=0$ requires only $180^\circ-2\varphi$ of rotation. The two shift movements at the begin and at the end of the scan ensure a $180^\circ$ sampling range in the whole field of measurement. 
}
\label{ShiftRays}
\end{figure}

The proposed C--arm trajectory requires a robotic motorization of the orbital axis of the C and of ideally two orthogonal axes in the C plane. The trajectory consists of three parts: An orbital rotation of the C around a defined iso--center and two translational movements parallel to the detector plane at the begin and at the end of the rotation. As proposed in \cite{tita:07} the orbital rotation can also be performed using a non isocentric C. Combining these three segments to one trajectory enables for the acquisition of a completely sampled dataset using only  $\text{180}^\circ$ minus fan--angle of rotation. Figure \ref{ShiftImages} shows these three parts of the trajectory examplarily. The main idea is, that a pure translation of x--ray source and detector creates an angular change of the ray intersecting a fixed point inside the volume of interest. This effect is caused by the fan--beam geometry which describes mostly all x--ray C--arm systems used today. Figure \ref{ShortRays} and Figure \ref{ShiftRays} visualize the change of the angular sampling range during a scan for the short scan (Figure \ref{ShortRays}) and for the rotate--plus--shift scan (Figure \ref{ShiftRays}). Here one can clearly see, that after a rotation of $180^\circ$ minus fan angle one point at the border ($x=0$ in Figure \ref{ShortRays} and \ref{ShiftRays}) of the field of view (FOV) is already fully sampled with $180^\circ$ angular coverage. However, at the opposite side of the FOV ($x=2\RM$, with $\RM$ being the radius of the field of view, in Figure \ref{ShortRays} and \ref{ShiftRays}) the angular coverage is $4\varphi$ lower (with $\varphi$ being the half fan angle). Here the  short scan method simply extends the rotation by $4\varphi$ which leads to a rotation range of $180^\circ+2\varphi$. The rotate--plus--shift trajectory instead makes use of the fact that a C rotation of $180^\circ$ minus fan angle yields exactly $180^\circ$ angular coverage in one point ($x=0$) of the volume. If these rays are shifted from one border ($x=0$) of the FOV to the other ($x=2\RM$) all points inside the FOV will be sampled in a range of $180^\circ$. Figure \ref{ShiftRays} shows these single steps in detail. As the diagram in the lower row shows, using a rotation of exactly $180^\circ-2\varphi$ leads to a sampling without redundant rays, which might be beneficial in terms of dose usage. 

The proposed trajectory is not limited to circular rotations. The rotational part of the trajectory can be performed on an arbitrary path around the patient which might be optimized for collision protection or for imaging peripheral parts of the body (shoulder, hip). Also the shift movement does not have to be performed on a straight line, the only important point is, that the ray at $x=0$ is translated to $x=2\RM$.

\subsection{Image reconstruction}
%For reconstructing the circle plus shift data an adapted version of the FDK (Feldkamp-Davis-Kress) algorithm  \cite{feldkamp:84,wiesent:00} can be used. Before %backprojection projection data which are truncated due to the shift motion, are extrapolated using a standard approach before convolution \cite{sourbelle:05}. 
For reconstructing the circle plus shift data in this study a standard SART algorithm \cite{andersen:84} was used. We perfomed three iterations on each dataset using a subset size of 10 projections. 

For a clinical use it will be of interest to develop a reconstruction based on filtered backprojection. Especially if the angular range of the rotational part is larger than $\text{180}^\circ$ minus fan--angle a dedicated redundancy weight must be applied prior to convolution and backprojection. Defining such a reconstruction approach including a correct weighting function will be part of future investigations and is out of the scope of this study.
\section{Results}
To verify the method proposed both simulated data and experimental data were used. 
\subsection{Simulation study}
\begin{figure}[t!]
\centering\includegraphics[width=0.8\textwidth]{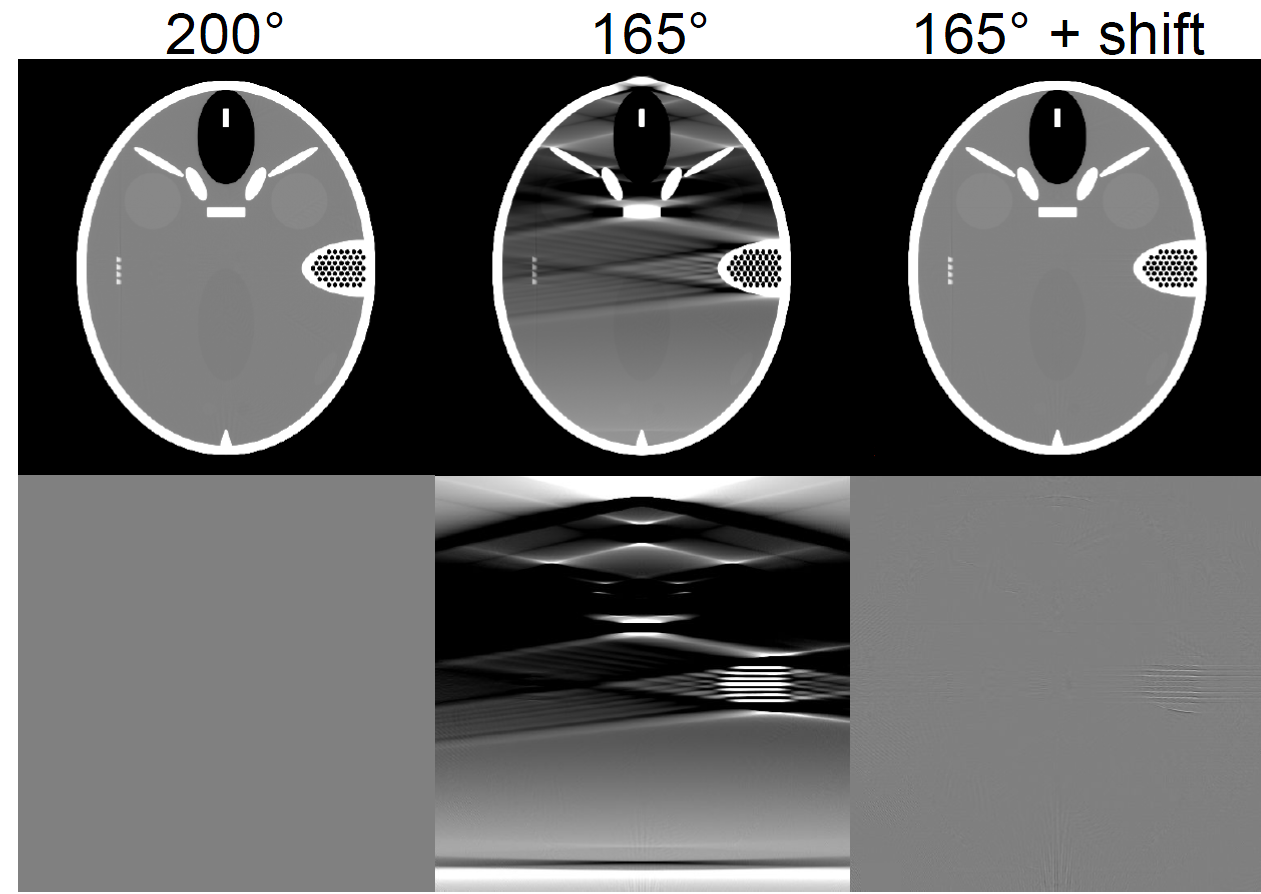}
\caption{Simulation of the FORBILD head phantom. From left to right: $200^\circ$ short scan, $165^\circ$ limited angle scan, $165^\circ+\text{shift}$ scan. The lower row shows difference images to the short scan reconstruction. Note the strong artifacts of the limited angle scan compared to the rotate--plus--shift scan. Both scans are simulated with only $165^\circ$ rotation. The structures in the difference image between the short scan and the rotate--plus--shift scan are due to differences of spatial sampling. All images are windowed $C=0~\text{HU}$ , $W=200~\text{HU}$.
}
\label{Simulation}
\end{figure}
For a first evaluation simulations using the FORBILD head phantom were performed. Figure \ref{Simulation} shows reconstructions of a $200^\circ$ short scan, a $165^\circ$ limited angle scan and a $165^\circ+\text{shift}$ scan. As expected the short scan and the rotate--plus--shift scan offer comparable image quality due to the full angular sampling of the whole field of view. The limited angle scan instead is strongly affected by limited angle artifacts. The data were simulated in fan--beam geometry using a source to detector distance of $1080~\text{mm}$ and a detector size of $300~\text{mm}$. 
\subsection{Experimental C--arm data}
\begin{figure}[h!]
\centering\includegraphics[width=\textwidth]{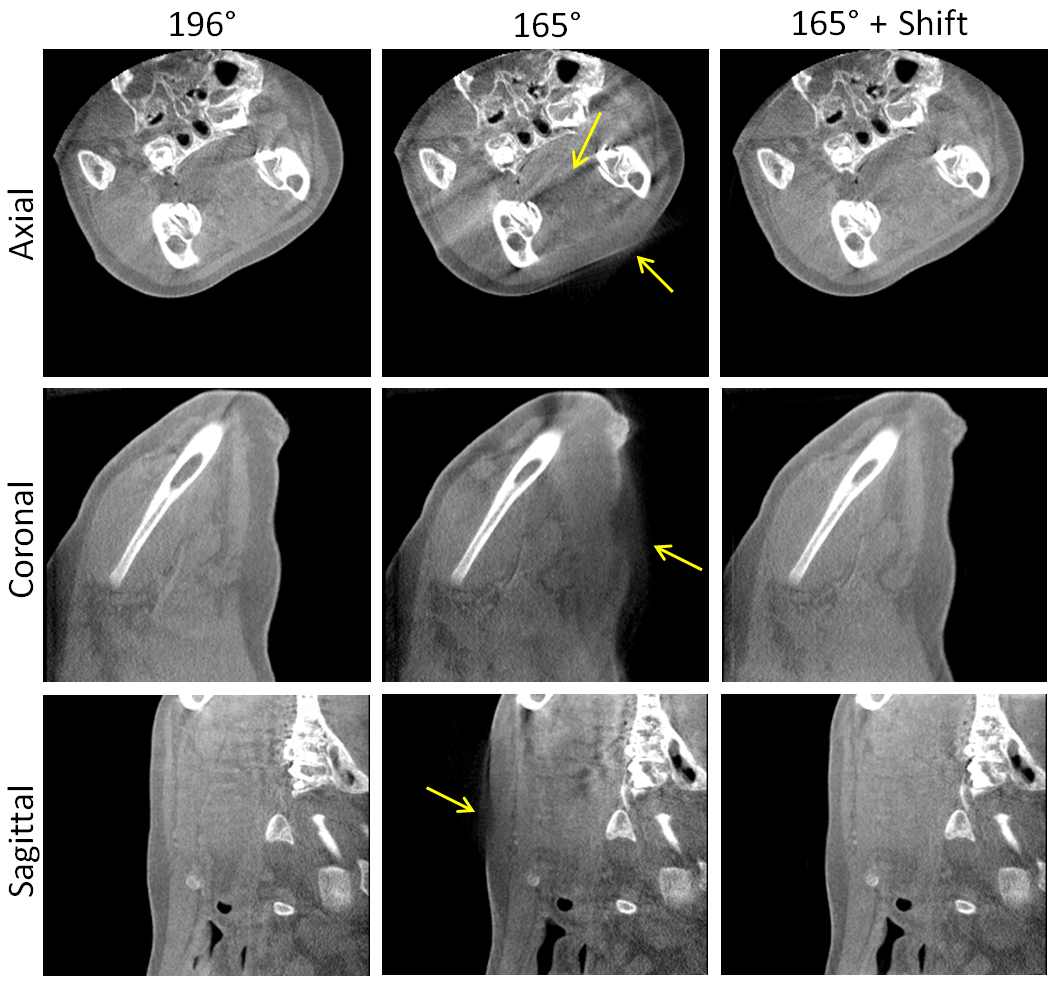}
\caption{The first dataset shows a circular short scan over $\text{196}^\circ$, which is used as reference data set. The second one shows the reconstruction of data acquired only over a limited angular range of $165^\circ$. The third reconstruction shows data acquired using the proposed rotate--plus--shift trajectory. Note the strong artifacts of the $165^\circ$ scan (yellow arrows). Both the short scan and the rotate--plus--shift scan offer comparable image quality. All images are windowed $C=0~\text{HU}$ , $W=1000~\text{HU}$
}
\label{PigHead}
\end{figure}

For an experimental validation C--arm scans of a living pig were acquired. All measurements were conducted at the German Cancer Research Center (DKFZ, Heidelberg, Germany) and approved by the local authorities (Regierungspr\"asidium Karlsruhe, 35-9185.81/G-118/13). The C--arm used for the acquisition is a mobile C--arm prototype with a flat detector (Varian Paxscan 3030+) offering $\text{196}^\circ$ of orbital rotation. Additionally two orthogonal axis in the C plane are motorized. This allows for the parallel acqusition of both the circular short scan trajectory and the rotate--plus--shift trajectory. The source to detector distance of the C--arm is about $1080~\text{mm}$, the detector size is $300~\text{mm}\times300~\text{mm}$. This leads to a fan angle of about $15.5^\circ$. Based on the previous considerations only about $165^\circ$ of orbital rotation are sufficient for a complete data set when using the rotate--plus--shift trajectory. 
Figure \ref{PigHead} shows the reconstructions of three datasets, a $\text{196}^\circ$ short scan, a $\text{165}^\circ$ limited angle scan and a $\text{165}^\circ$ plus shift scan. For both the short scan and the rotate--plus--shift trajectory scan time was about $40~\text{s}$. The short scan and the shift scan were acquired using $512$ projections, the $\text{165}^\circ$ limited angle scan using $431$ projections. The tube voltage was set to $80~\text{kV}$, the tube current time product of the short scan and the rotate--plus--shift scan was $235~\text{mAs}$. All parts of the projections which include rays outside the field of measurement during the shift motion were collimated using an adaptive collimator. Also here the limited angle scan shows strong artifacts while the short scan and the rotate--plus--shift scan yield comparable image quality. Note, that the short scan is acquired at about $8\%$ higher dose ($4\varphi/(2\cdot180)=0.0861$, ratio of redundant rays shown in Figure \ref{ShiftRays}) than the rotate--plus--shift scan due to redundant rays.  
\section{Discussion and Conclusion}
\begin{figure}[h!]
\centering\includegraphics[width=\textwidth]{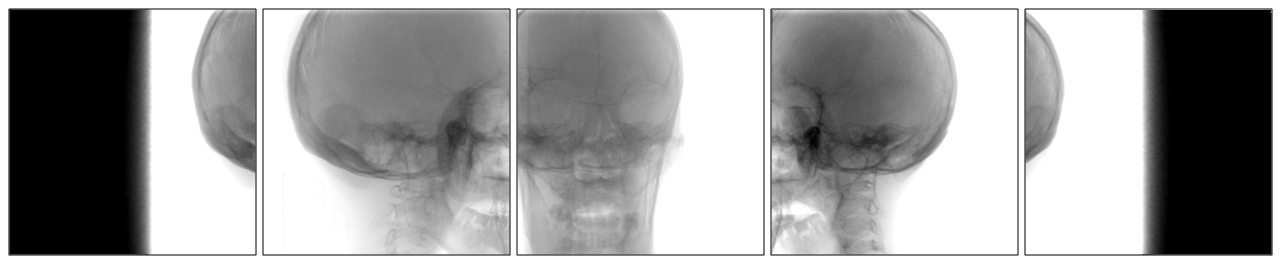}
\caption{Exemplary projection images of an antropomorph head phantom using the proposed rotate--plus--shift trajectory are shown. The first and the last image were acquired during the shift motion. Parts of the projection which cover regions outside the field of view are collimated.
}
\label{Shiftscan}
\end{figure}
In this article we presented a proof of concept study of a novel C--arm trajectory which overcomes limitations in the design of currently used C--arm devices. First experimental results using a C--arm prototype show that the rotate--plus--shift trajectory yields 3D datasets of comparable image quality compared to datasets acquired using a circular short--scan trajectory which is assumed as the gold standard for 3D C--arm imaging today. Regarding patient dose the rotate--plus--shift scan enables a sampling of the field of measurement without redundant rays, which leads to lower acqusition dose compared to the short scan. All parts of the projections during the shift motion which include rays outside the field of measurement should be collimated during the scan (Figure \ref{Shiftscan}). If an adaptive collimation is also used for the short scan, radiation dose will be at comparable levels. 

Further investigations, which are required for a clinical implementation, are the adaption of filtered backprojection--based algorithms including correct redundancy weighting. 

The rotate--plus--shift method enables for integrating full 3D functionality into a C- arm device without any loss of handling and usability for 2D imaging. We expect that the transition of this method into clinical routine will lead to a much broader use of intraoperative 3D imaging in a wide field of clinical applications.

\section*{Acknowledgements}
This work was supported by DFG grant No. KA 1678/11.
%\begin{thebibliography}{10}

%\end{thebibliography}
\bibliographystyle{ieeetr}
\bibliography{MK}
%\nocite{Wagner:89}

\end{document}